# Two-stage SQUID amplifier with bias current re-use

**Mikko Kiviranta**
VTT, Tietotie 3, 02150 Espoo, Finland

E-mail: Mikko.Kiviranta@vtt.fi

**Abstract.** We have designed and fabricated an integrated two-stage SQUID amplifier, requiring only one bias line and one flux setpoint line. From the biasing viewpoint the two stages are connected in series while from the signal propagation viewpoint the stages are cascaded. A proof-of principle demonstration at $T$ = 4.2 K is presented.

## 1. Introduction

The long-standing VTT-SRON collaboration [1] has sought to develop SQUID amplification chains for Frequency Domain Multiplexed (FDM) readout [2] of Transition Edge Sensors (TESes) [3]. The work has been motivated by the X-IFU instrument [4] on board the ATHENA space observatory and the SAFARI instrument [5] on board the recently cancelled SPICA space observatory.

In the rudimentary form of FDM the TESes are ac voltage biased with a set of carrier frequencies, and TES currents are summed with a T-junction, the total current being read out with a SQUID ammeter. A side benefit of the ac bias is that the required low-impedance bias source for the TESes can be reactive, either inductive ([1] Fig. 5) or capacitive ([6] Fig. 5), and hence dissipationless. Because the T-junction combiner does not provide isolation between the input branches (unlike eg. [7]), the SQUID ammeter should ideally represent a perfect short. In practice the input inductance $L_{IN}$ of the SQUID ammeter should be very small to prevent pixel-to-pixel crosstalk.

The small $L_{IN}$ causes two kinds of difficulties. First, the required energy resolution of the ammeter, expressed as $\varepsilon = \frac{1}{2} L_{IN} i_n^2$ in terms of the input-referred current noise $i_N$, becomes demanding. Because the $L_{IN}$ restriction becomes tighter when the number of multiplexable TES pixels increase, the energy constraint can be considered similar to the 'noise penalty'[1] [8] in the Time Domain Multiplexing (TDM). Second, because small $L_{IN}$ samples only a fraction of the available signal power generated by the TESes, the power gain of the amplification chain needs to be larger. In practice two or three cascaded

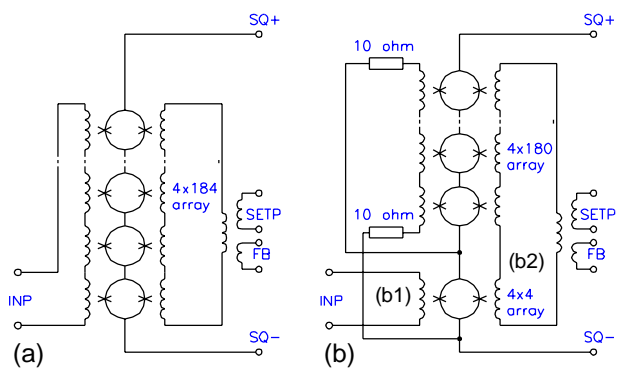

Figure 1: Simplified internal schematic (a) of the F5p -type booster and (b) of the F6p -type two-stage amplifier derived from the F5.

SQUID stages are needed to obtain sufficient signal power to drive the room-temperature Low Noise Amplifier (LNA) over the (nominally) 2 K to 300 K cable.

The originally conceived VTT-SRON amplification chain consisted of two cascaded stages, the front end (FE) SQUID and the booster (AMP) SQUID [9], and formed the basis of the system-level design of the X-IFU and SAFARI instruments. Two new constraints emerged later, causing the two-stage amplifier to be short in power gain: linearity of the amplification chain and the length $l$ of the 2 K to 300 K cable. As large length as $l \geq 12$ m was required by the cryogenic design of the SPICA telescope in a certain phase of development. Two solutions were pursued to obtain more power gain: (i) increase of the critical current density $J_C$ of the Josephson junctions [10] in the two-stage cascade, and (ii) introduction of the third intermediate SQUID. The third stage should ideally not require more bias wires, to avoid a significant revision of the system-level design. The amplifier

---

[1] Note that the FDM penalty can in principle be alleviated by a proper power combiner [7], and the TDM penalty by pulsed TES bias [1].

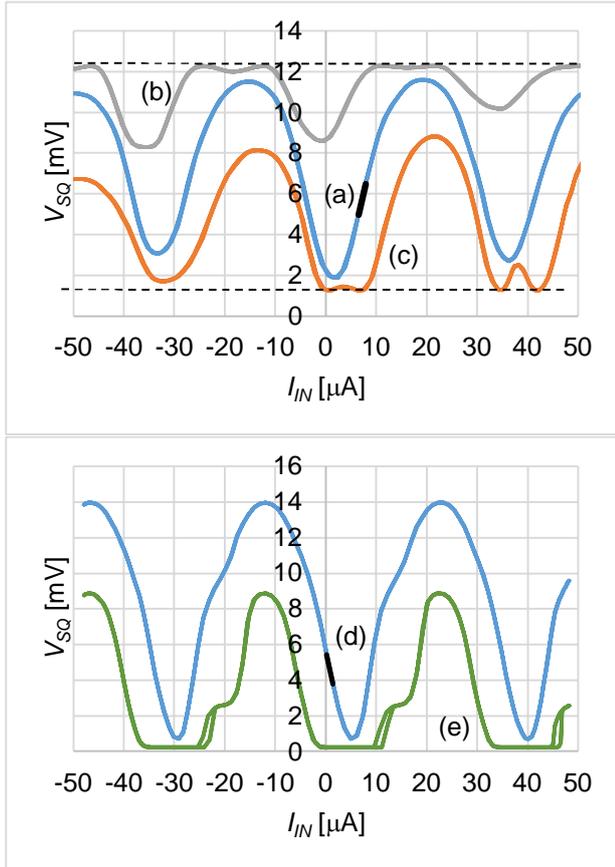

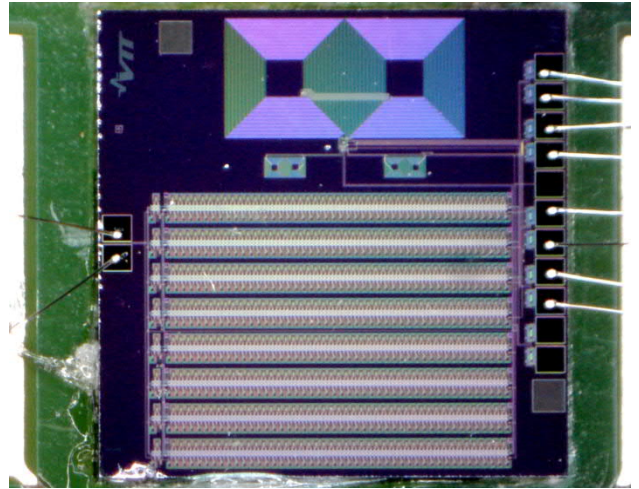

Figure 3: Microphotograph of the F6-type two-stage amplifier on a 4 x 4 mm chip.

Figure 2: (a-c) Measured flux-to-voltage characteristics of the F6-type two-stage amplifier, excited through the input coil. Dashed lines indicate the voltage swing of the 180x4 second stage, through which the flux response of the 4x4 first stage is amplified. Flux setpoint coil has been used (a) to center the 1st stage response, and (b, c) to make the the response clip at the edges of the 2nd stage monotonous flux range.
(d, e) Measured flux-to-voltage characteristic of the F5-type booster at two different bias current values. The input coil is connected to 2 x 4.99 kΩ cryogenic resistors generating the coil current.

described here is a proof-of-principle demonstrator for the latter approach.

## 2. Device design and tests

### 2.1 The F5 booster

The two-stage amplifier described here is a modification of the 184-series 4-parallel SQUID array, used as the booster in the TES experiments at SRON [11]. The SQUID cells are of gradiometric construction with $L_{SQ}$ = 40 pH loop inductance, containing Josephson junctions with the nominal critical current $I_C$ = 25 µA and $R_S$ = 4.5 Ω shunt resistors. The cells are equipped with a 2-turn input coil, which leads to $M^{-1}$ = 35 µA/$\Phi_0$ periodicity. We estimate an input inductance of $L_{IN,AMP}$ = 150 nH. Our designation for the 184 x 4 array design is 'F5', with the design naming schema detailed out in the Table 1 in the Appendix. Owing to their 'narrowline' design, the F5 boosters can be operated in the earth field without magnetic shielding. The devices described here are of the generation patterned with projection lithography [12] and designated as 'F5p'.

Characteristics registered from an F5p sample are shown in Fig. 2 (d) and (e). When the input coil is open or connected to a high-resistance source, the device tends to develop Enpuku plateaus [13, 14], which disappear when a low-resistance source is connected to the input. Bandwidth and flux noise are shown in Fig. 4 (d) and (c), measured in liquid helium with $l$ = 1 m twisted pair BeCu loom [15] as the wiring. A homemade, actively terminate-able, fully differential SiGe amplifier [10, 16] was used as the LNA. We estimate that the spectral spikes in the 1 - 10 MHz region of the trace c of the Fig. 4 are due to electromagnetic emission of our data acquisition instrument. The apparent white noise floor of $\Phi_N$ = 45 n$\Phi_0$/Hz$^{1/2}$ includes a contribution by the LNA: the $u_N \approx$ 0.5 nV/ Hz$^{1/2}$ voltage noise divided by the booster gain $dV/d\Phi$ = -48 mV/$\Phi_0$ at the measurement setpoint. plus a contribution of the LNA current noise $i_N \approx$ 2 pA/ Hz$^{1/2}$ and Johnson noise of the wiring.

### 2.2 The F6 two-stage amplifier

As depicted in Fig.1, the two-stage amplifier is formed by dividing the 184-series 4-parallel booster SQUID array into 180-series 4-parallel and 4-series 4-parallel SQUID sections, sharing the same bias current. The voltage across the 4 x 4 section is sampled by 2 x 10 Ω on-chip resistors and fed to the input coil of the 180 x 4 section which acts as the second stage of the cascade. The biasing arrangement where constant current biased sections are connected in series is an electrostatic dual of ordinary semiconductor amplifier ICs where constant voltage biased sections are connected in parallel. The division of the array into 180 x 4 and 4 x 4 sections is visible in the left side of the chip microphotograph Fig. 3. We designate the device type as 'F6p'.

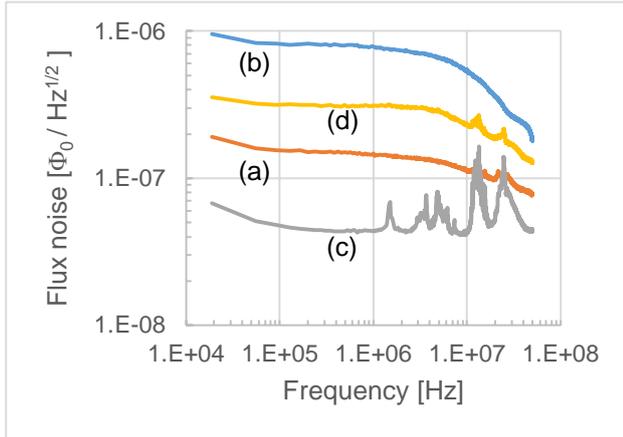

Figure 4: (a) Flux noise and (b) frequency response of the F6-type two-stage amplifier, measured at the flux setpoint emphasised in the trace a of the Fig. 2. (c) Flux noise and (d) frequency response of the F5-type booster, measured at the flux setpoint emphasized in the trace d of the Fig. 2. Frequency responses are obtained with pseudorandom noise excitation.

The flux characteristics of the 4 x 4 first stage, as recorded through the second stage, are shown as trace a in Fig. 2. White flux noise of $\Phi_N = 160$ n$\Phi_0$/Hz$^{1/2}$ is shown in the trace a of the Fig. 4, which appears to be close to the true flux noise of the first stage. The amplifier shows $dV/d\Phi = 38$ mV/$\Phi_0$ at that setpoint. We estimate $L_{IN,INT} \approx 3$ nH when interconnect parasitics are neglected.

We remark that the 1$^{st}$ stage flux response in Fig. 2 does not precisely repeat from period to period. We attribute this to the fact that the setpoint coils of the 4 x 4 first stage and the 180 x 4 second stage are not driven from a stiff current source, but from the self-inductance $L_{TRA} \approx 150$ nH of the summing transformer visible in the Fig. 1b. Hence the flux setpoint of the 180 x 4 second stage does not remain fixed, but is affected by the input coil via the parasitic b1-to-b2 coupling in the Fig. 1. The motivation for the presence of the summing transformer is to provide three simultaneously useable flux-adjusting coils for the F5 device: flux setpoint, current-sampling local feedback and voltage sampling local feedback. Only one feedback is shown in Fig. 1.

## 3. Concluding discussion

We have demonstrated the general feasibility of adding a third intermediate SQUID stage to the 2-stage SQUID tandem, without a need for additional bias lines. Adding a 7:1 transformer in front of the F6-type amplifier would raise the transresistance $R_{TR} = (dV/d\Phi)/M^{-1} \approx 1.4$ k$\Omega$ shown by the unmodified F5-type booster into $R_{TR} \approx 7.5$ k$\Omega$ range, while both devices would show comparable $L_{IN} \approx 150$ nH input inductance. This achieves the goal of the proof-of-principle demonstration.

Additionally, we find conceivable use for the unmodified F6-type or a comparable low $L_{IN}$ device in the (non-multiplexed) readout of very fast TES detectors, such as the photon number resolving optical TESes [17].

## Acknowledgement

The SQUID array design and fabrication was supported by the E-SQUID project, grant no. 262947 of the European Community, 7th framework programme (FP7/2007-2013). The testing has been supported by the Academy of Finland through the Centre of Exellence for Quantum Technology.

## Appendix

**Table 1.** SQUID generations for TES multiplexing at VTT

| Devices available | Device designations | Features |
|---|---|---|
| 05/2009 | A*: FDM front-end, narrowline<br>B*: FDM front-end, wideline<br>C*: FDM boosters, wideline<br>D*: boosters, narrowline | 100 mm wafer, contact lithography, Pd resistors, pillar junctions [9]. |
| 07/2013 | D*: legacy boosters<br>E*: supercond. transformers<br>F*: boosters<br>G*: front-ends<br>H*: novel devices | 100 mm wafer, contact lithography, TiW resistors, pillar junctions |
| 05/2014 | Code domain and time domain multiplexers, non-letter designated | 150 mm wafer, projection lithography, TiW resistors, pillar junctions [12] |
| 08/2014 | D*p: legacy boosters<br>F*p: legacy boosters<br>G*p: legacy front-ends<br>J*: A-type front-ends, modified for anti-coincidence detector readout | 150 mm wafer, projection lithography, TiW resistors, pillar junctions [12] |
| 06/2017 | K*: front-ends, AntiCo readouts<br>L*: boosters | 150 mm wafer, projection lithography, TiW resistors, pillar junctions |
| 10/2019 | M*: front-ends, AntiCo readouts<br>N*: boosters | 150 mm wafer, projection lithography, TiW resistors, SWAPS junctions [10] |